google.c%% ****** Start of file apstemplate.tex ****** %
\documentclass[%
superscriptaddress,
preprint,
amsmath,
amssymb,
aps,
prb,
]{revtex4}

\usepackage{graphicx}% Include figure files
\usepackage{dcolumn}% Align table columns on decimal point
\usepackage{bm}% bold math
\usepackage{hyperref}% add hypertext capabilities
\begin{document}

\preprint{}

\title{Hysteretic Magnetotransport in SmB$_{6}$ at Low Magnetic Fields}

\author{Y. S. Eo}
	\email{eohyung@umich.edu}
	\affiliation{University of Michigan, Dept.~of Physics, Ann Arbor, Michigan 48109-1040, USA}
\author{S. Wolgast}
	\affiliation{University of Michigan, Dept.~of Physics, Ann Arbor, Michigan 48109-1040, USA}
\author{T. \"{O}zt\"urk}
	\affiliation{University of Michigan, Dept.~of Physics, Ann Arbor, Michigan 48109-1040, USA}
	\affiliation{Sel\c{c}uk University, Dept.~of Physics, Konya, Turkey, 42075}
\author{G. Li}
	\affiliation{University of Michigan, Dept.~of Physics, Ann Arbor, Michigan 48109-1040, USA}
\author{Z. Xiang}
	\affiliation{University of Michigan, Dept.~of Physics, Ann Arbor, Michigan 48109-1040, USA}
	\affiliation{University of Science and Technology of China, Hefei National Laboratory for Physical Science at Microscale and Department of Physics, Hefei Anhui 230026, China}
\author{C. Tinsman}
	\affiliation{University of Michigan, Dept.~of Physics, Ann Arbor, Michigan 48109-1040, USA}
\author{T. Asaba}
	\affiliation{University of Michigan, Dept.~of Physics, Ann Arbor, Michigan 48109-1040, USA}
\author{F. Yu}
	\affiliation{University of Michigan, Dept.~of Physics, Ann Arbor, Michigan 48109-1040, USA}
\author{B. Lawson}
	\affiliation{University of Michigan, Dept.~of Physics, Ann Arbor, Michigan 48109-1040, USA}
\author{J. W. Allen}
	\affiliation{University of Michigan, Dept.~of Physics, Ann Arbor, Michigan 48109-1040, USA}
\author{K. Sun}
	\affiliation{University of Michigan, Dept.~of Physics, Ann Arbor, Michigan 48109-1040, USA}
\author{L. Li}
	\affiliation{University of Michigan, Dept.~of Physics, Ann Arbor, Michigan 48109-1040, USA}
\author{\c{C}. Kurdak}  
	\affiliation{University of Michigan, Dept.~of Physics, Ann Arbor, Michigan 48109-1040, USA}

\author{D.-J. Kim}
	\affiliation{University of California at Irvine, Dept.~of Physics and Astronomy, Irvine, California 92697, USA}
\author{Z. Fisk}
	\affiliation{University of California at Irvine, Dept.~of Physics and Astronomy, Irvine, California 92697, USA}
 
\date{\today}

\begin{abstract}

Utilizing Corbino disc structures, we have examined the magnetic field response of resistivity for the surface states of SmB$_{6}$ on different crystalline surfaces at low temperatures. Our results reveal a hysteretic behavior whose magnitude depends on the magnetic field sweep rate and temperature. Although this feature becomes smaller when the field sweep is slower, a complete elimination or a saturation is not observed in our slowest sweep-rate measurements, which is much slower than a typical magnetotransport trace. These observations cannot be explained by quantum interference corrections such as weak anti-localization. Instead, they are consistent with behaviors of glassy surface magnetic ordering, whose magnetic origin is most likely from samarium oxide (Sm$_{2}$O$_{3}$) forming on the surface during exposure to ambient conditions.   

\end{abstract}

%\pacs{73.25.+i, 73.20.-r}%

\maketitle

\section{\label{sec:Intro}Introduction}

Following recent work done on SmB$_{6}$, topological Kondo insulators (TKI) have emerged as a fascinating area of study where the fields of strongly correlated materials and topological insulators (TI) meet. SmB$_{6}$ is a long-studied mixed-valent insulator in which the insulating behavior arises from the hybridization between the localized 4$f$ electrons and the delocalized 5$d$ electrons at low temperatures \cite{Menth, Allen}. Recent theories suggest that SmB$_{6}$ is a strong 3D topological insulator caused by band inversion occurring at the high symmetry points \cite{Dzero, Takimoto, Dzero2, Alexandrov}. Within this picture, the bulk is expected to be an insulator and the surfaces to be a gapless metal at cryogenic temperatures. Since these surface states are topologically protected, they are expected to be robust against non-magnetic impurities and exhibit a helical spin structure in $k$-space. 

Experimentally, these expected robust metallic surface states and insulating bulk behavior were first verified by electrical transport measurements \cite{Wolgast, Kim}. Hybridization gap and metallic surface formation have been studied by a wide range of spectroscopic measurements including angle resolved photo emission spectroscopy (ARPES) \cite{Miyazaki, NXu, jjiang, Neupane, Denlinger1, Denlinger2}, point-contact spectroscopy \cite{ZhangJP}, and scanning tunneling spectroscopy \cite{Yee, Rossler, Ruan} experiments. These hybridization gap values are reported slightly differently, but are roughly in agreement at $\sim20$ meV.  

More recent theoretical reports suggest that there are three Dirac-like Fermi pockets at the surface \cite{Tran, FengLu, Alexandrov}. For the (001) crystalline surface, theory predicts that one of the pockets surround the $\Gamma$ point, and the other two surround the X points. The Dirac points inside these pockets are protected by time-reversal symmetry. Meanwhile, further theoretical work predicts that the two Dirac points located along the X-$\Gamma$ of the (011) surface are protected by crystalline mirror symmetry \cite{MYe}. Although the small energy scales involved in SmB$_{6}$ makes the APRES measurements particularly challenging, some of the high resolution ARPES measurements have revealed the odd number of Fermi pockets at the (001) crystalline surface \cite{NXu, Neupane}. There is even a report of the helical spin structure by spin-resolved ARPES measurements \cite{Xu_SpinARPES}. Angular-dependent magneto-torque measurements also show signs of multiple 2D pockets from the de-Haas van Alphen (dHvA) oscillations \cite{GLi}. The extrapolation of the Landau levels of the Fermi pockets to the infinite magnetic field limit shows a Dirac-like half-integer behavior. However, the size of the Fermi pockets measured by dHvA are not in complete agreement with the ARPES results. 

Because SmB$_{6}$ is a bulk insulator, magnetotransport measurements can also be used to probe the 2D surface states. In particular, quantum oscillations and weak anti-localization (WAL) are expected. Normally for a two-dimensional electron gas (2DEG), Shubnikov-de Haas oscillations (SdH) can be used to extract the carrier density ($n_{2D}$) and mobility ($\mu_{2D}$). However, so far there is no convincing evidence of SdH oscillations up to 45 T \cite{GLi, Wolgast2}, suggesting that the surface has a low mobility (can be order of 100 cm$^{2}$/V$\cdot$ sec). On the other hand, WAL is a quantum interference effect which provides a signature of the helical spin structure of a 3D TI \cite{HikamiLarkinNagoaka, HasanRMP, AndoReview}. Consistent with the expectation of a TI, WAL are recently reported on SmB$_{6}$ by two groups: Xia's group and Paglione's group \cite{SThomasWAL, NakajimaJP}. 

In this paper, we study the surface transport of SmB$_{6}$ using a Corbino disc structure. The resistivity value of a Corbino disc corresponds to the longitudinal component of conductivity ($\sigma_{xx}$) in a Hall bar structure with an additional magnetic field dependence by the geometrical shape which can be neglected at the examined magnetic field regime. The key advantage of this setup is that it allows us to study the magnetic response to resistivity for individual surfaces. Recently, Corbino structures have been utilized to extract the carrier density and mobility of SmB$_{6}$ by applying high magnetic fields and ionic gating \cite{Wolgast2, SyersJP}. Our resistivity data at low magnetic fields exhibit hysteretic behaviors which become stronger at faster magnetic field sweep rates. At a fixed sweep rate, the feature caused by this hysteresis resembles WAL. However, the strong sweep-rate dependence suggests that this feature is not caused by a quantum interference effect, but rather caused by a magnetic origin. We attribute this dynamically slow hysteresis to a glassy magnetic ordering behavior possibly by the native samarium oxide (Sm$_{2}$O$_{3}$) formed on the surface after exposure to ambient air.  

\section{\label{sec:Methods}Experimental Methods}

Single-crystal SmB$_{6}$ samples were grown by the Al flux method. The samples were thinned either using a coarse SiC grit manually or a lapping machine that automates the process using Al$_{2}$O$_{3}$ slurry. Each sample was thinned so that at least one Corbino disc pattern could fit either on the (001) or (011) crystalline surface. Typical final surface areas were in a few mm$^{2}$, and the final thickness ranged from $\sim$ $500$ $\mu$m up to 1 mm. Then, we polished the samples using a fine grit or slurry until the surface quality was good enough to perform photolithography. 

On top of the polished SmB$_{6}$ surface, we patterned a Corbino disc with an inner (outer) diameter of $300$ $\mu$m ($500$ $\mu$m) using standard photolithography. A descum process was performed on the sample surface using oxygen plasma. Next, $5/150$ nm  Ti/Au was evaporated, and the active region was lifted-off. Au or Al wires were bonded using wire bonding, and silver paint was occasionally added on top of the contacts for better adhesion. For most of our samples, two wires for each source and drain were bonded so that the resistance of the wires can be neglected by four-terminal measurements. One of our samples with a complete Corbino disc with contacts is shown in the inset of Figure~\ref{fig:HighFieldMR}. All of our contacts were Ohmic at both room and cryogenic temperatures ($4.2$ K or below). Electrical measurements were taken by standard techniques using a lock-in amplifier, and occasionally adding a pre-amplifier with a bridge circuit to achieve clearer signals. The time constant that determines the low pass filter bandwidth of the lock-in amplifier was set short enough ($\tau$ = 1 sec) so that even at our fastest magnetic field sweep rates (32 mT/sec) the associated time delay is not significant. The excitation current ($I$ = 10$^{-7}$ to 10$^{-6}$ A) was sufficiently small that the measured resistance did not depend on the current or frequency. Cryogenic temperature measurements were performed in a $^3$He cryostat system, and a $^3$He/$^4$He dilution refrigerator using a superconducting magnetic with a bipolar magnetic power supply, whose current polarity switching occurs at $B$ $\neq$ 0 T. To examine varying field-angle dependence, additional measurements were performed at the National High Magnetic Field Laboratory.  

\section{\label{sec:Results}Measurements of Magnetotransport at Low Magnetic Fields}

\begin{figure}
\includegraphics{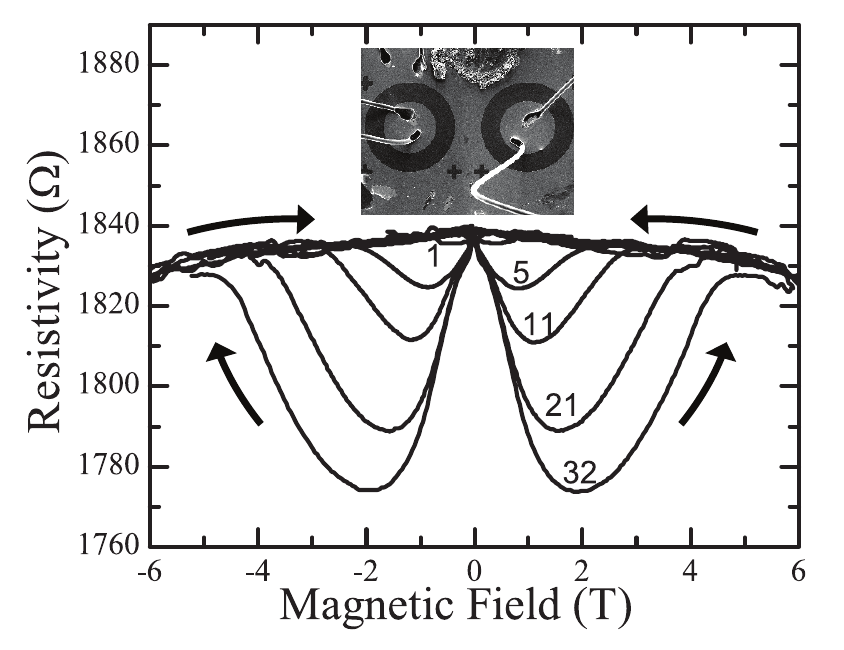}
\caption{Response of resistivity of the Corbino disc samples to the magnetic field at different sweep rates below 6 T at $0.3$ K. The numbers shown close to each curve are the magnetic field sweep-rate magnitude in units of mT/sec. The inset on top of trace shows an example of a Corbino disc sample image prepared on a polished SmB$_{6}$ surface.}
\label{fig:HighFieldMR}
\end{figure}

The response of resistivity to the magnetic field shows slow dynamical hysteretic behaviors.  Specifically, the resistivity is dependent on the history of the magnetic field and its sweep rate. For a systematic study, we start from a large magnetic field value ($-B_{max}$ to $+B_{max}$) and measure resistivity sweeping in both directions at different field sweep rates ($dB/dt$). Figure~\ref{fig:HighFieldMR} shows typical resistivity traces of one of our Corbino disc samples at different sweep rates. This dynamical hysteretic behavior was observed in all of our samples. Following the arrows in this figure, while sweeping the magnetic field from $-6$ T ($-B_{max}$) until $0$ T, the resistivity does not show any strong features. However, continuing from $0$ to $+6$ T ($+B_{max}$), a noticeable dip occurs. The resistivity first starts to decrease and reaches to some minimum value. Then, the resistivity starts to return to its path as the magnetic field is further increased. Changing the sweep direction, and sweeping from $+6$ T ($+B_{max}$) to $0$ T, this dip does not appear. Continuing from 0 to $-6$ T ($-B_{max}$), the strong dip appears again. As a result, the two strong dips appear symmetrically on each polarity of the magnetic field. By increasing the magnetic field sweep rate, the magnitude of these dips becomes larger. Typically, these dips appear at magnetic fields smaller than $\pm$5 T. 

We note that trivial heating from the sample or from an external source (e.g. the magnet power supply) cannot explain this behavior. First, Joule heating of the sample cannot be the case. By changing the current through the sample by an order of magnitude we did not observe a change in the hysteretic behavior. Also, Joule heating of the sample is orders of magnitude smaller than the cooling power of our cryogenic system. Second, inductive heating by eddy currents cannot be the case. Inductive heating depends on the magnetic field sweep rate, but is independent of the sweep direction. Since inductive heating is constant throughout a fixed-sweep rate, if inductive heating causes the resistivity change of the sample, this change can only be monotonic and non-reproducible as sweeping several cycles. However, our data have two non-monotonic dips which are reproducible at a constant sweep rate and temperature. Also, comparing to the cooling power at $0.3$ K, the magnitude of inductive heating is orders of magnitude smaller. Finally, if a single polar power supply is used for the superconducting magnet, it can cause a dip in resistivity as it switches the circuit at zero magnetic field. For this reason, we used a bipolar magnetic power supply for which the switching event (B $\neq$ 0 T) was identified, and we confirmed that the dips are independent from this event. 
\begin{figure}
\includegraphics{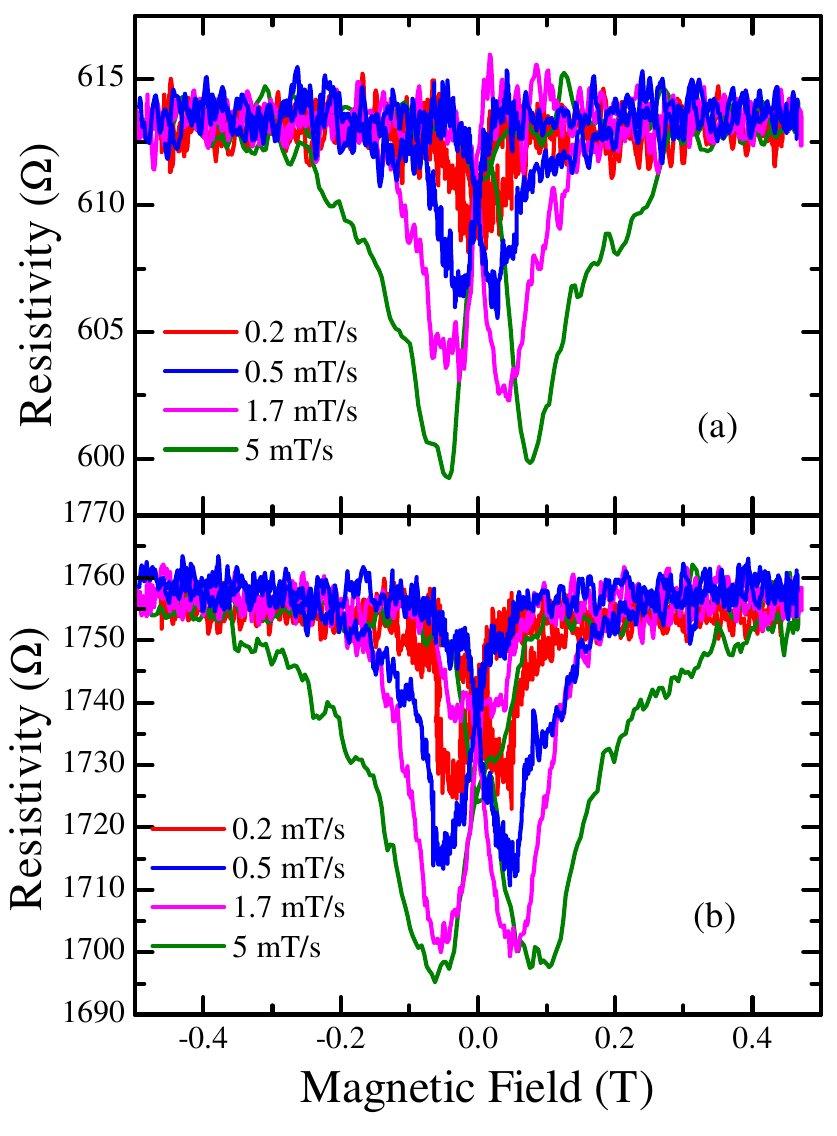}
\caption{(Color) Response of resistivity of the Corbino disc samples to the magnetic field at different sweep rates below 1 T at 80 mK. (a) (001) sample (b) (011) sample.}
\label{fig:LowFieldMR}
\end{figure}

\begin{figure}
\includegraphics{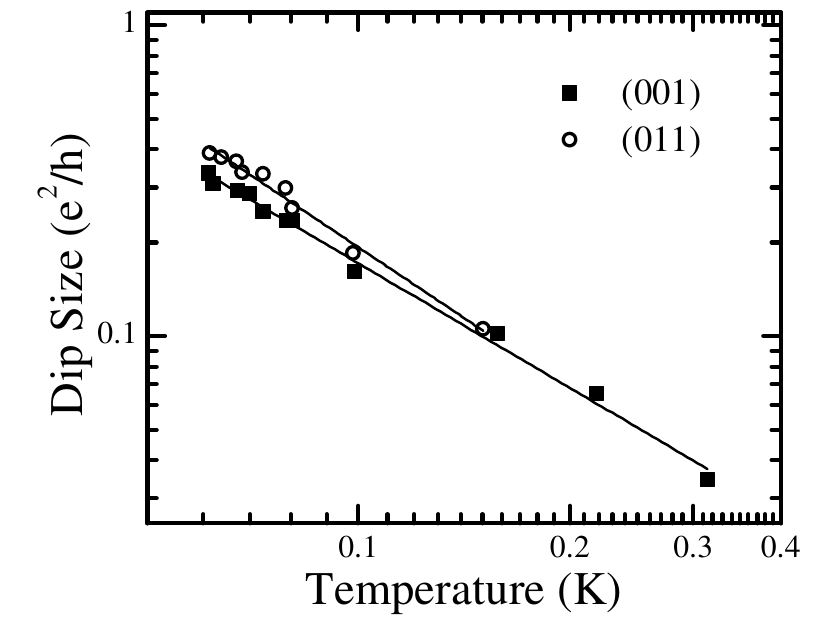}
\caption{Magnitude of the dips (in conductivity) as a function of temperature at magnet field sweep rate $0.167$ mT/s.}
\label{fig:DipTemp}
\end{figure}

\begin{figure}
\includegraphics{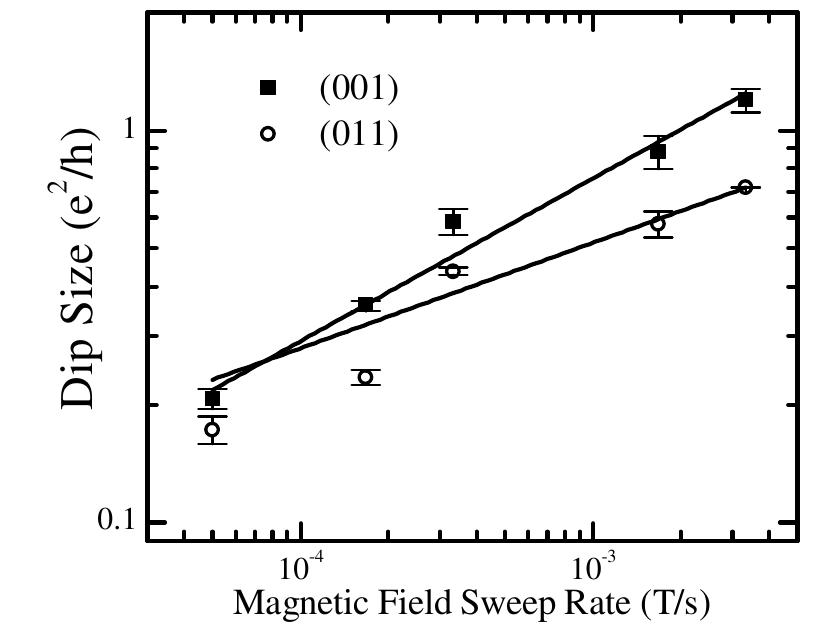}
\caption{Magnitude of the dips (in conductivity) as a function of magnet field sweep rate at 80 mK.}
\label{fig:DipSweep}
\end{figure}

\begin{figure}
\includegraphics{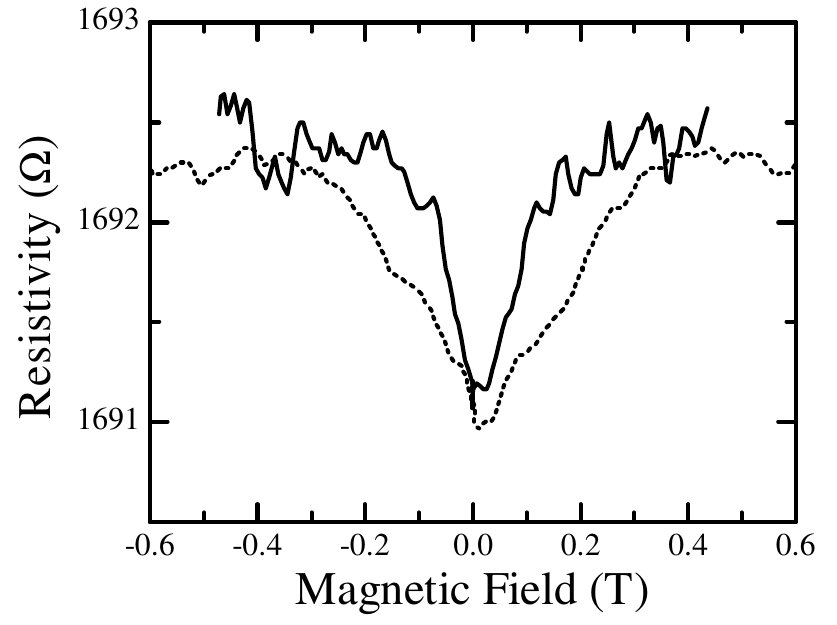}
\caption{Response of resistivity of the Corbino disc samples to the magnetic field comparing at different angles of magnetic field at $0.3$ K. Solid curve is the magnetic field perpendicular to the transport surface. Dotted curve is magnetic field parallel to the transport surface. The minimal points are shifted to $B$ = 0 T for direct comparison.}
\label{fig:Angle}
\end{figure}

Previously, WAL has been reported\cite{SThomasWAL, NakajimaJP} within this magnetic field range. The sweep-rate dependent dynamic dips that we observe as shown in Figure~\ref{fig:HighFieldMR} are not caused by WAL. For the WAL case, the magnetic field only breaks the phase of the electrons traveling a closed loop by static impurities, and this phase does not depend on $dB/dt$. In a further attempt to observe WAL, that is, a dip in resistivity which is non-dynamical and non-hysteretical, we measured resistivity at extremely slow magnetic field sweep rates and at lower temperatures for two samples, each prepared in different crystalline directions. As a result, the two samples show additional hysteretic features at lower magnetic field ranges (within $\pm$ 1 T). As shown in Figure~\ref{fig:LowFieldMR}, these features were systematically studied as described above (Figure~\ref{fig:HighFieldMR}) at lower temperatures (down to 60 mK). We observe dynamical hysteretic features with two symmetric dips similar to those in Figure~\ref{fig:HighFieldMR}. The hysteretic features are smaller, and the positions of the minima appear at a lower field range, but the qualitative magnetic field response remains the same. If one were to fix on a particular magnetic field sweep rate, the data shows similarities to WAL. As shown in Figure~\ref{fig:DipTemp}, when converting the magnitude of the dips ($\Delta R$) to change in conductivity ($\Delta\sigma$), the sizes are on the order of typical WAL peak magnitudes ($\sim$0.1 e$^{2}$/h). Also, similarly to WAL, $\Delta\sigma$ increases as the temperature is lowered. However, the magnetic field response must be static for WAL. Although the magnitude of the dips decreases at slower sweep rates, we did not observe any sign of the dip magnitude to saturate (become non-dynamic). The magnitude of the dips as a function of magnetic field sweep rate for both samples is shown in Figure~\ref{fig:DipSweep}. Even at the slowest measurements ($dB/dt$ = 5$\times$10$^{-5}$ T), which takes more than 5 hours to sweep 1 T, the magnitude of the dips continues to shrink. In addition to this measurement, we took angle dependent magnetic field measurements that also indicate that this feature is not WAL. WAL can only be observed as a function of the perpendicular magnetic field component\cite{HikamiLarkinNagoaka}. As shown in Figure~\ref{fig:Angle}, however, the dips also appear in parallel (in-plane) magnetic fields, and this dip widens very slowly than what we expect from a typical WAL feature as the field is rotated from the perpendicular to the parallel direction \cite{HikamiLarkinNagoaka}. 

To the best of our knowledge, there is no theory that can provide a quantitative description of our data. In the next section, we provide the possible origin and a qualitative description of our hysteretic data.  

\section{\label{sec:Discuss}Possible Origin of the Dynamical Magnetotransport Feature}

Our hysteretic magnetotransport data suggest that magnetic properties play an important role. Since the SmB$_{6}$ crystal is nonmagnetic\cite{Menth}, we do not expect the measured magnetic properties to arise from the bulk. Furthermore, the hysteretic behavior is only observed at low temperatures when the surface dominates the transport. We suggest that magnetic properties may arise from the Sm$^{3+}$ present in the native samarium sesquioxide (Sm$_{2}$O$_{3}$) that is formed on the surface of the samples.

Recent X-ray magnetic circular dichroism and X-ray absorption spectroscopy spectra show that Sm$^{3+}$ is dominant on the surface with a net magnetic moment\cite{Phelan}. In addition, hard X-ray photoelectron spectroscopy (HAXPES) shows a weak oxygen signal of a polished and then etched SmB$_{6}$ sample\cite{HaoTjeng}. These results imply Sm$_{2}$O$_{3}$ oxide is formed when the surface of SmB$_{6}$ is exposed to air at ambient conditions\cite{HaoTjeng}. The native Sm$_{2}$O$_{3}$ formed on the SmB$_{6}$ surface is expected to be disordered. 

The long time scales associated with the strong magnetic field sweep-rate dependence in our data suggest this magnetic system is glassy. In a spin glass system, the relaxation time of magnetization can be extremely long\cite{BinderSpinGlassRMP}. Therefore, when an external magnetic field is applied, we expect the magnetization to depend on the magnetic field sweep rate even at very slow rates. In addition, the total magnetization of a spin glass system exhibits a hysteresis loop, so the area of the hysteresis loop depends on the magnetic field sweep rate. Theoretically, the hysteresis area becomes larger at faster sweep rates, and at lower temperatures\cite{Sariyer}. Similarly to our measurements, when the magnetic field is swept from a large to a small field strength, the magnetization decreases. Since the magnetization follows a hysteresis loop, zero magnetization appears past zero magnetic field (at coercive field) symmetrically with respect to the sweep direction. 

We expect that our resistivity data's response to the magnetic field is a manifestation to this magnetization property of a glassy system. In a scenario where the resistivity decreases when the magnetization decreases, our data's magnetic field history, sweeping direction, sweep rate, and temperature dependence are consistent with the magnetization of the spin glass features explained above. We expect the positions of the dip minima occurs when the magnetization is zero (at the coercive fields of a hysteresis curve). In addition, for high enough magnetic fields, the spin glass phase typically breaks into an ordinary magnetically ordered phase \cite{AlmeidaThouless}, and this can explain why our hysteretic behavior becomes weaker at high magnetic fields ($B$ $>$ 5 T). 

Within the context of transport experiments, understanding the coupling between the glassy system and the conducting surface states is a critical task. The magnitude and sign of the magnetic response to resistivity are non-universal. They depend on the microscopic details of the glassy system, as we discuss the possibilities below. In a simple picture where domains are formed in the system, the domain boundary between them serves as a scatterer, and this will cause the resistivity to increase. However, according to the single domain boundary model by Gorkom et al., the domain boundary can instead result in a decrease of the resistivity, depending on the relaxation times and distributions of the spin orientations in the magnetic system\cite{Gorkom}. In another scenario, for a surface of a time-reversal invariant topological insulator, stronger magnetization would destroy the time-reversal symmetry protection, and thus the resistivity would be enhanced. If the total magnetization is reduced, time-reversal symmetry breaking will be less, and thus the surface will be less resistive. Meanwhile, a more specific model has been proposed for ferromagnetically doped systems in which each magnetic domain boundary can be treated as a quantized 1D conducting channel, decreasing the overall resistivity\cite{Nomura}. 

At low temperatures when surface transport dominates, there is a logarithmic increase in resistivity as the temperature is lowered, and a negative magnetoresistance at high magnetic fields up to 35 T\cite{Wolgast2}, both of which are signatures of Kondo scattering. Typically, Kondo scattering and spin glass phase do not coexist, since the Kondo cloud formed by the conduction carriers and the local spin of the magnetic impurity will quench any magnetic ordering at low temperatures. Also, the Kondo effect is usually expected in the dilute limit of magnetic impurity concentration. Since we expect that Sm$_{2}$O$_{3}$ on the surface of SmB$_{6}$ is not dilute\cite{Phelan, HaoTjeng}, a sophisticated picture is required to account for both the spin glass and Kondo scattering\cite{Wolgast2}. 

For this picture, we consider a disordered Kondo lattice system\cite{Theumann, Magalhaes} forming a spin glass system. This Kondo lattice is dense, and is formed by conduction carriers from the SmB$_{6}$ surface interacting with a disordered dense array of localized moments from the Sm$_{2}$O$_{3}$. If we first consider an ordered Kondo lattice, as the temperature is lowered from high temperatures, the resistivity rises logarithmically as the magnetic ordering becomes quenched by a Kondo cloud formation, where the spin scattering between the localized $f$-electrons and the conduction electrons inside it increases. As the temperature is lowered further, the resistivity drops since the effect of coherence between the lattice sites (Bloch's theorem) dominates, and the magnetic moment is quenched\cite{Coleman}. Now, when the Kondo lattice system is disordered, this downturn due to coherence can be averted, and we can expect a remnant magnetic moment to exist in the system. There are examples of heavy fermion systems that show suppression of this downturn by introducing even a small doping amount\cite{Scoboria, Amakai}. In addition to this resistivity behavior, glassy states induced by disorder are also known\cite{Tang, Sullow} in heavy fermion systems. The persisting disordered magnetic moments can lead to the frustration of exchange interactions; therefore, in some cases, this disordered Kondo lattice system can lead to a spin glass state\cite{Theumann, Magalhaes}. Here we note that the recent HAXPES results indicate that the oxide thickness is $20\pm5$ \AA\cite{HaoTjeng, HaoTjengPersonal}, which this thickness can be regarded as a 2D system\cite{Mitric, Adachi}. Long-range spin order for a 2D system is typically more vulnerable to disorder effects, compared to a 3D system, and thus has a stronger tendency towards a glassy ground state. For example, for the 2D Ising order, it is well known that even an infinitesimal random field is sufficient to destroy the long-range spin order and can turn the system into a 2D glassy state\cite{Cardy, ImryMa}. For a topological surface, a theory work considering magnetic adatoms on top\cite{Abanin} predicts that this system can be in a spin glass state when frustrated in-plane exchange interactions are larger than out-of-plane interactions, a condition favored by 2D magnetic layers. 

However, even if this is the case, we point out that the actual system that becomes glassy can be much more complicated. In particular, the origin of disorder may differ from the previously studied heavy fermion spin glass cases. Previously studied disordered Kondo lattice models assume that RKKY interaction strengths are randomly distributed\cite{Theumann, Magalhaes}. However, a simple model of a topological insulator surface involving two magnetic impurities implies that RKKY interactions might or might not be able to make the Sm$_{2}$O$_{3}$ on top of the SmB$_{6}$ surface glassy\cite{Liu}. In this model, RKKY interactions between two magnetic impurities can only be in a ferromagnetic ordering state if the Fermi energy lies near the Dirac point ($a_{0} k_{F} \approx 0$). Using the $k_{F}$ values of SmB$_{6}$ from high magnetic field and recent ionic gated Corbino measurements, the position of the Fermi energy is near or slightly above the Dirac point ($a_{0} k_{F} \approx 1$). From these results, it is inconclusive that ferromgnetic ordering can be destroyed and become a glassy state$\cite{Wolgast2, SyersJP}$. Instead, since Sm$_{2}$O$_{3}$ is an insulator\cite{Pan}, random superexchange interactions, which are independent of $k_{F}$, may be also responsible for the glassy state. An ordered Sm$_{2}$O$_{3}$ is expected to be anti-ferromagnetic ordering at low temperatures ($\theta_{W}\approx$ $-4$ K)\cite{Mitric}. Since oxides can change from an anti-ferromagnetic to a ferromagnetic ordering state as the angle of the chemical bonds change, we expect that Sm$_{2}$O$_{3}$ can be glassy when they are distributed randomly on a surface with different bond angles\cite{Akamatsu, Kanamori, Mitric2}. An incorporation of this aspect to theory may be required for further understanding.  

\section{\label{sec:level1}Conclusion}

In conclusion, we have investigated magnetotransport of SmB$_{6}$ at low magnetic fields using Corbino disc structures. All of our samples revealed a dip of resistivity which were magnetic field sweep-rate dependent. Although these features become smaller in magnitude at slower sweep rates, the magnitude is still clearly visible at our slowest measurements. These features are most likely due to extrinsic magnetic impurity scattering by the naturally formed samarium oxide (Sm$_{2}$O$_{3}$), which exhibits a glassy magnetic ordering. Thus, the behavior of the dip is inconsistent with WAL, and to the extent permitted by the dip at the slowest sweep rates, we do not observe WAL.

A topological insulator with no bulk contribution can potentially be an ideal building block for realizing Majorana Fermions, and spintronics devices\cite{Wilczek, Fu, Yokoyama}. If eliminating the hysteretic magnetotransport behavior, which is suspected to be magnetic, the surface of SmB$_{6}$ may be a strong candidate of this building block. For preventing the native samarium oxide formation on the SmB$_{6}$ surface, growing a heterostructure or a cap layer on top of the SmB$_{6}$ surface may be a possible solution. 

\begin{acknowledgments}
We wish to acknowledge Kyunghoon Lee for his assistance with wirebonding the Corbino contacts, and Juniar Lucien for polishing crystal surfaces. This work was supported by the National Science Foundation grants ECCS-1307744, DMR-1006500, DMR-1441965, DMR-0801253, the Department of Energy award de-sc0008110, the Scientific and Technological Research Council of Turkey (TUBITAK), China Scholarship Council, and the National Basic Research Program of China (973 Program, Grant No. 2012CB922002). Device fabrication was performed in part at the Lurie Nanofabrication Facility, a member of the National Nanotechnology Infrastructure Network, which is supported by the National Science Foundation. The high-field experiments were performed at the National High Magnetic Field Laboratory, which is supported by NSF Cooperative Agreement No. DMR-084173, by the Cooperative Agreement No. DMR-084173, by the State of Florida, and by the DOE.
\end{acknowledgments}

\bibliography{HysteresisReference}

\end{document}